\documentclass[preprint, showkeys, 
superscriptaddress,
amsmath,amssymb,
aps,
prb,
]{revtex4-2}
\usepackage{pgfplots}
\usepackage{pgfplotstable}
\usepackage{booktabs}
\usepackage{array}
\usepackage{colortbl}
\usepackage{graphicx}
\usepackage{dcolumn}
\usepackage{bm}

\usepackage{xcolor} 
\usepackage{physics} 
\pgfplotsset{compat=1.18}
\begin{document}

\preprint{APS/123-QED}

\title{Effects of boundary conditions on quantum nanoresonators: decoherence-free subspaces}

\author{Humberto C. F. Lemos}
 \email[Corresponding author:]{humbertolemos@ufsj.edu.br}
\affiliation{%
 Departamento de Estatística, Física e Matemática - CAP - Universidade Federal de São João del-Rei, Brazil.
}%

\author{Thiago Cordeiro}
\email{thiago.cordeiro69@educacao.mg.gov.br}
\affiliation{Secretaria de Estado de Educação de Minas Gerais. Belo Horizonte, MG, Brazil.}%

\author{Adelcio C. Oliveira}%
 \email{adelcio@ufsj.edu.br}
\affiliation{%
 Departamento de Estatística, Física e Matemática - CAP - Universidade Federal de São João del-Rei, Brazil.
}%

\date{\today}

\begin{abstract}
The Euler-Bernoulli beam model has been studied classically and semi-classically. The semi-classical quantization is done in an analogous way to the quantization of the electromagnetic field, and we found an effect that is similar to the Casimir effect, which is the photonic Casimir effect. The Casimir force, by unit area, is proportional to the first mode energy divided by the volume of the beam.
For the hinged-hinged boundary condition, degenerate states were found. These degenerate pairs form decoherence-free subspaces for dispersive thermal reservoirs. For other boundary conditions, there are also subspaces with lower decoherence rates, which occur for quasi-degenerate states.

\end{abstract}
\maketitle
 \newpage 

\section{Introduction} \label{sec:intro}

The first quantum revolution was about the building blocks of the computer, but its working functionality is as classical as that of the first mechanical computer. Now we face a second quantum revolution when the algorithm works according to quantum logic. To achieve a quantum computational advantage, we must utilize quantum properties of the system that have no classical analog. This is known as a quantum resource, and the study of this area is referred to as quantum resource theory \cite{Brandao}. For one-mode states, there are two resources, the negativity of the Wigner function \cite{NEG} and the Squeezing \cite{Sinatra}. While negativity is present in many applications and can be considered, in some cases, as an entanglement witness \cite{Negativity_entanglement}, non-local states signature \cite{Nonlocality_Neg,Nonlocality_Neg2}, and it is strongly correlated with purity. On the other hand, squeezing is an important issue to the improvement of metrology \cite{Quantum_metrology}. For many-mode states, the entanglement is the most used resource \cite{Yin}. An important challenge, if not the most, is to keep the quantum coherence \cite{QuantumCoherence}; thus, it is important to know what makes the system behave classically. A quantum system behaves classically due to a combination of three factors: large actions, interaction with the environment, and measurement disturbance \cite{OLIVEIRA2012,Oliveira2014}.
To avoid these troubles, small actions are usually considered, and many different strategies of measurement, such as non-demolition measurements \cite{QND_Measurements} and weak interactions \cite{POVM}. The interaction with the environment depends on the particular experimental realization, with large advances in many areas \cite{Deco1,Deco2,Deco3}.

Nanomechanical beams are mechanical structures that can be fabricated \cite{Poot2008}. They serve as useful models for nanotubes \cite{Nanotubes} and microtubules \cite{Quant_microtubulo,Ruan,CIVALEK20112053}, and have various applications \cite{resonator_application}. There is also significant interest in the quantum properties of these systems, with numerous studies exploring their quantum effects and applications \cite{POOT,Quantviga,Quantviga_rev,Quant_opt_mec}, as well as investigations into philosophical questions such as consciousness \cite{Penrose}. In the classical domain, these mechanical systems exhibit rich dynamics, including chaotic behavior \cite{Awrejcewicz,Zhou, Pellicano, Chen, Zhang,Wei, Liu,Norouzi,Cao,CaosV1,CaosV2} and stochastic dynamics \cite{VigaTherm}.

Due to the large number of particles, modeling a nanoresonator or a nanobeam is not a simple task. Classical models limit their treatment as a continuous medium, and discrete aspects of the material's lattice may be relevant. Among the approaches is atomic modeling, which essentially involves solving the Schrödinger equation for a system of \(N\) atoms. Atomic modeling is usually divided into molecular dynamics, Monte Carlo, density functional theory, tight bonding molecular dynamics, local density, Morse potential function model, and ab initio approaches \cite{RAFIEE2014435}. 

In this work, we will investigate the continuum model \cite{RAFIEE2014435}, based on the Euler-Bernoulli beam model. Continuum models have already been studied in several other works \cite{Yakobson1996,Andras2008,RAFIEE2014435}; here, our interest is in semiclassical modeling and how boundary conditions affect the dynamic behavior of the quantum nanobeam.

The classical and consequently, the quantum solutions of the nanobeams are strongly affected by the contour conditions, and while boundary conditions are extensively investigated in classical domains, there are few attempts to capture its effects in the quantum domain \cite{Jiang2012}. In our model, different boundary conditions imply in significantly different sensitivity of environmental action, suggesting the existence of decoherence-free subspaces \cite{Decoherence_free_exp,Decoherence_free_sub1,Decoherence_free_sub2}. The work is organized as follows: in \ref{sec:classical} we briefly review the theoretical modeling of classical beams. In section \ref{sec:quant}, we investigate the quantum version for the clamped-clamped, clamped-hinged, and clamped-free boundaries. In section \ref{sec:Dechoerence_free} we investigate the possible existence of decoherence-free subspaces for phase damping reservoir \cite{Phasereservoir,HilbertSpace}. In section \ref{sec:Conclusions} we present our conclusions.

\section{Classical beam modeling} \label{sec:classical}

\begin{figure}
    \centering
    \includegraphics[width=0.85\linewidth]{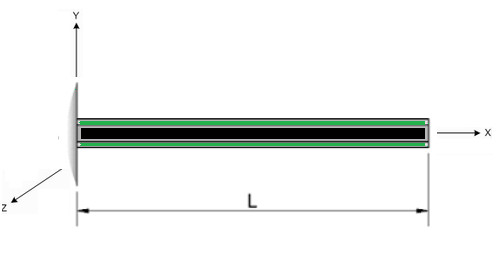}
    \caption{Euler-Bernoulli beam schematic model.}
    \label{fig:beam}
\end{figure}

We begin by studying the behavior of an elastic beam under small transverse vibrations, subjected to different loads and under different boundary conditions. Consider a beam of length \( L \) as illustrated in figure \ref{fig:beam}: the reader may notice that the reference frame is set such that the \( x \)-axis runs along the span of the beam. 

We briefly summarize the main results from the Euler-Bernoulli beam theory, with detailed discussions available in standard textbooks, such as \cite{balachandran2018}. We assume that the beam's displacements are confined to the \( xz \)-plane and denote them by \( w(x,t) \), where positive \( w \) indicates downward displacement. The potential energy of the beam is given by
\begin{equation} \label{eq:classicbeam_potential}
    U(t) = \frac{1}{2} \int_{0}^{L} EI \left( \frac{\partial^{2} w}{\partial x^{2}} \right)^{2} dx ,
\end{equation}
where \( E \) is the Young's modulus of the material, and \( I \) represents the area moment of inertia of the beam's cross-section about the \( y \)-axis. The kinetic energy of the beam is
\begin{equation} \label{eq:classicbeam_kinetic}
    T(t) = \frac{1}{2} \int_{0}^{L} \rho A \left( \frac{\partial w}{\partial t} \right)^{2} dx ,
\end{equation}
where \( \rho \) is the mass density of the beam and \( A \) is the cross-sectional area of the beam about the \( y \)-axis.

Applying the energies derived above, the Lagrangian mechanics formalism yields a general equation for the beam's behavior, as detailed in \cite{balachandran2018}. The analysis initially considers a homogeneous beam with a uniform cross-section along its length. It is further assumed that no axial load acts along the beam and that the beam does not rest on an elastic foundation. Under these conditions, the vertical deflection \( w \) as a function of horizontal position \( x \) and time \( t \) is governed by the following fourth-order linear non-homogeneous partial differential equation (PDE):
\begin{equation} \label{eq:classicbeam_dyn}
    EI\frac{\partial^4 w}{\partial x^4}+\rho A \frac{\partial^2 w}{\partial t^2}=f(x,t),
\end{equation}
for \( 0 < x < L\), where \( f \) represents an external transversal loading per unit length of the beam. To find the general solutions for this equation, we first consider the homogeneous case where \( f = 0 \). This homogeneous PDE can be solved using the method of separation of variables. We assume the solution takes the form \( w(x,t) = \xi (x) \tau (t) \), with the precise form of the spatial solution depending on the boundary conditions. There are many known boundary conditions for this problem \cite{balachandran2018}. For our initial study, we consider the hinged-hinged case, leading to the boundary conditions
\begin{equation} \label{eq:classicbeam_BCs}
    w = 0 \ , \quad \mbox{and } \ \frac{\partial^{2} w}{\partial x^{2}} = 0,
\end{equation}

at both boundaries of the beam, namely at \( x = 0 \) and \( x = L \). Alternative boundary conditions are discussed in a subsequent section.

The partial differential equation (PDE) \eqref{eq:classicbeam_dyn} can be separated into two linear ordinary differential equations (ODEs):
\begin{subequations} \label{eq:classicbeam_ODEs}
    \begin{align}
        \xi^{(4)}(x) - \lambda^{4} \xi(x) = 0, \label{eq:classicbeam_ODEs_x} \\
        \ddot{\tau}(t) + \displaystyle{\frac{\rho A}{EI}} \, \lambda^{4}  \tau(t) = 0.  \label{eq:classicbeam_ODEs_t}
    \end{align}
\end{subequations}
By applying the boundary conditions \eqref{eq:classicbeam_BCs} to Eq. \eqref{eq:classicbeam_ODEs_x}, we can determine the eigenvalues \( \lambda \), which correspond to the normal modes of the beam
\begin{equation} \label{eq:classicbeam_normalmodes}
    \lambda_{k} = \frac{k \pi}{L} , \quad k = 1, 2, \ldots ,
\end{equation}
with the spatial solutions given by the eigenfunctions
\begin{equation} \label{eq:classicbeam_spatial}
    \xi _{k} (x) = \sin (\lambda_{k} x) ,
\end{equation}
for \( k \geq 1 \). It is important to note that different boundary conditions yield distinct normal modes. The hinged-hinged case is analytically tractable and serves as a useful example, although it is challenging to implement in a nanometer-scale experimental apparatus due to the analytical determination of frequencies. The vibrational modes of the beam are examined in greater detail in subsection \ref{boundary}, where the implications of this choice are discussed. Under the boundary conditions \eqref{eq:classicbeam_BCs}, the beam equation \eqref{eq:classicbeam_dyn} admits an infinite number of solutions, resulting in the well-known general solution:
\begin{equation} \label{eq:classicbeam_generalsolution}
    w (x,t) = \sum_{k \geq 1} w_{k} (x,t) = \sum_{k \geq 1} \tau_{k} (t) \sin (\lambda_{k} x) .
\end{equation}
Once the eigenvalues are determined, solving Eq. \eqref{eq:classicbeam_ODEs_t} for each \( k \geq 1 \) becomes straightforward, as it reduces to the equation of a harmonic oscillator with a natural frequency given by
\begin{equation} \label{eq:classicbeam_frequencies}
    \omega_{k} = \sqrt{\frac{\rho A}{EI}} \, \lambda_{k}^{2} = \sqrt{\frac{\rho A}{EI}} \, \left( \frac{k \pi}{L} \right)^{2}.
\end{equation}
The solution for Eq. \eqref{eq:classicbeam_ODEs_t} is widely known. If initial conditions for the beam were provided, the complete classical solution \( w(x,t) \) could be determined. However, we are not interested in solving the classic beam problem, as this has already been thoroughly explored. Instead, we will proceed with the quantization of these harmonic oscillators to study nanoresonators.

\section{Quantization: semiclassical approach} \label{sec:quant}

We now proceed to the next step: the quantization of the beam. Specifically, we use the classical solution for the spatial part of the problem, given by the eigenfunctions \eqref{eq:classicbeam_spatial}. As discussed in Section \ref{sec:classical}, the temporal counterpart is described by an infinite set of harmonic oscillators, each corresponding to a different mode \( k \geq 1 \) in Eq. \eqref{eq:classicbeam_ODEs_t}. To study the quantum behavior of the system, we take the quantum version of these oscillators. Here, we consider that the beam is composed of countless particles, thousands or more. These nano-beams can, in fact, be monitored in real time, with their collective position and momentum measured almost constantly. Thus, we assume that the continuous monitoring regime has been reached \cite{Oliveira2014}, and that Newtonian dynamics is valid. However, continuous monitoring does not affect the spectrum of the Hamiltonian, so the temporal part is treated by a quantum in order to obtain part of the energy levels.

The quantum harmonic oscillator is a well-established problem, making its quantization straightforward. However, to validate our approach, we first evaluate the total energy of the beam. Substituting the general solution \eqref{eq:classicbeam_generalsolution} into the potential energy expression \eqref{eq:classicbeam_potential}, and using the orthogonality of the eigenfunctions \eqref{eq:classicbeam_spatial}, the potential energy of the beam is found to be
\[
    U(t) = \sum_{k \geq 1} \frac{E I L}{4} \, \lambda_{k}^4 \tau_{k}^{2} (t) ,
\]
Similarly, the kinetic energy is given by
\[
    T(t) = \sum_{k \geq 1} \frac{\rho A L}{4} \, \dot{\tau}_{k}^{2} (t) .
\]
Thus, the total energy of the beam, expressed as the Hamiltonian, is
\begin{equation} \label{eq:classicbeam_Hamiltonian}
    \mathcal{H} (t) = \sum_{k \geq 1} \frac{1}{2} \, \frac{\rho A L}{2} \, \dot{\tau}_{k}^{2} (t) + \frac{1}{2} \, \frac{E I L}{2} \, \lambda_{k}^4 \tau_{k}^{2} (t) .
\end{equation}
This Hamiltonian is precisely that of an infinite set of decoupled harmonic oscillators, one for each mode \( k \geq 1\). We define the generalized coordinates as \( q_{k} (t) = \tau_{k} (t) \), therefore the generalized velocities are \( \dot{q}_{k} (t) = \dot{\tau}_{k} (t) \). The first term in the summation, representing the kinetic energy, can be recognized as that of a harmonic oscillator with mass \( m = \rho A L /2 \). Such identification is two-fold: firstly, the potential energy term can be rewritten as
\[
    U(t) = \sum_{k \geq 1} \frac{1}{2} \, \frac{\rho A L}{2} \frac{E I}{\rho A} \, \lambda_{k}^4 \tau_{k}^{2} (t) =  \sum_{k \geq 1} \frac{1}{2} \, m \omega_{k}^{2} q_{k} (t) ,
\]
which corresponds to the potential energy of a harmonic oscillator with mass \( m \) and frequency \( \omega_{k} \) given by \eqref{eq:classicbeam_frequencies}. Additionally, if we define the generalized momenta as \( p_{k} (t) = (\dot{\tau}_{k} (t) / m) \), the Hamiltonian can be rewritten as
\begin{equation} \label{eq:classicbeam_Hamiltonian_momentum}
    \mathcal{H} (t) = \sum_{k \geq 1} \frac{1}{2} \, \frac{p_{k}^{2} (t)}{m} + \frac{1}{2} \, m \omega_{k}^{2} q_{k} (t) .
\end{equation}

With this formulation, we are ready for quantization. The total energy \eqref{eq:classicbeam_Hamiltonian} is replaced by the Hamiltonian operator \( \hat{\mathcal{H}}\). Correspondingly, position and momentum are replaced by the operators \( \hat{q}_{k} \) and \( \hat{p}_{k} \), which satisfy the canonical commutation relations. We define the annihilation and creation operators, \( \hat{a}_{k} \) and \( \hat{a}_{k}^{\dagger} \), respectively, as
\begin{subequations} \label{eq:quantumbeam_a_adaga}
    \begin{align}
        \hat{a}_{k} &=  \sqrt{\frac{m \omega_{k}}{2 \hbar}} \left( \hat{q}_{k} + \frac{i}{m \omega_{k}} \hat{p}_{k} \right) ,  \label{eq:quantumbeam_a}
        \\
        \hat{a}_{k}^{\dagger} &=  \sqrt{\frac{m \omega_{k}}{2 \hbar}} \left( \hat{q}_{k} - \frac{i}{m \omega_{k}} \hat{p}_{k} \right) . \label{eq:quantumbeam_adaga}
    \end{align}
\end{subequations}
These operators satisfy the commutation relations
\begin{equation} \label{eq:quantumbeam_a_adaga_comutador}
    [\hat{a}_{k} , \hat{a}_{k'}] = 0 = [\hat{a}_{k}^{\dagger} , \hat{a}_{k'}^{\dagger}] , \quad [\hat{a}_{k} , \hat{a}_{k'}^{\dagger}] = \delta_{k, k'}.
\end{equation}
For each \( k \geq 1\), the number operator is \( \hat{N}_{k} = \hat{a}_{k}^{\dagger} \hat{a}_{k} \), and so the Hamiltonian is
\begin{equation} \label{eq:quantumbeam_Hamiltonian}
    \hat{\mathcal{H}} = \sum_{k \geq 1} \hat{\mathcal{H}}_{k} = \sum_{k \geq 1} \hbar \omega_{k} \left( \hat{N}_{k} + \frac{1}{2} \right) ,
\end{equation}
where \( \hat{\mathcal{H}}_{k} = \hbar \omega_{k} (\hat{N}_{k} + 1/2) \) is the Hamiltonian for each \( k \).

As expected, the number operator \( \hat{N}_{k} \) satisfies the commutation relations
\[
    [\hat{N}_{k} , \hat{a}_{k'}] = -\hat{a}_{k} \delta_{k, k'} , \quad [\hat{N}_{k} , \hat{a}_{k'}^{\dagger}] = \hat{a}_{k}^{\dagger} \delta_{k, k'} .
\]
Moreover, for each \( k \geq 1 \) and for each \( n \geq 0\), the state \( \ket{k, n} \) is an eigenvector of \( \hat{N}_{k} \), with eigenvalue \( \hat{N}_{k} \ket{k, n} = n \ket{k, n} \). Consequently, it is an eigenvector for \( \hat{\mathcal{H}}_{k} \), i.e. \( \hat{\mathcal{H}}_{k} \ket{k, n} = \varepsilon_{k,n} \ket{k, n} \), and its eigenenergy is given by
\begin{equation} \label{eq:quantumbeam_epsilon_kn}
    \varepsilon_{k, n} = \hbar \omega_{k} \left( n + \frac{1}{2} \right) = \frac{\hbar \pi^{2}}{L^{2}} \sqrt{\frac{\rho A}{EI}} \, k^{2} \left( n + \frac{1}{2} \right) .
\end{equation}
At this stage, the model is considered solved in a formal sense. However, we will now delve deeper into specific aspects to further analyze the implications of our findings.

\subsection{Renormalization}

Lets consider  \( k \geq 1 \) and  \( n \geq 0\), we can write $\varepsilon_{k,n}$ as 
\begin{equation} \label{epskn}
    \varepsilon_{k,n} = C k^{2} \left( n + \frac{1}{2} \right) ,
\end{equation}
and \( C \) is defined by \eqref{eq:quantumbeam_epsilon_kn}. The vacuum energy is  \( \varepsilon_{k, 0} = C k^{2}/2 \), thus considering all states, the sum $\sum_k \varepsilon_{k,n}$ is divergent. 

Now defining a ``re-normalized''mode energy as 
\begin{equation} \label{epskn0}
    \varepsilon_{k,n}^{(0)} = C k^{2} n .
\end{equation}
we have \( \varepsilon_{k, 0}^{(0)} = 0 \), also 
\[
    \varepsilon_{k,n} = \varepsilon_{k,n}^{(0)} + \frac{C k^{2}}{2} = \varepsilon_{k,n}^{(0)} + \varepsilon_{k, 0} .
\]

When we write \( \hat{\mathcal{H}}_{k} \) it is a compact way to say that 

\begin{equation}
     \hat{\mathcal{H}}_{k} = \underbrace{I \otimes \ldots \otimes I}_{(k-1) \mbox{ times}} \otimes \hat{\mathcal{H}}_{k} \otimes I \ldots ,
\end{equation}
and the state \( \ket{k,n} \) represents the state
\begin{equation}
    \ket{k,n} = \ket{1, 0} \otimes \ldots \ket{k-1, 0} \otimes \ket{k, n} \otimes \ket{k+1, 0} \otimes \ldots .
\end{equation}
The eigenvalue eigenvector equation given by \( \hat{\mathcal{H}} \ket{k, n} = E_{k, n} \ket{k, n} \), where the energy eigenvalues \( E_{k, n} \) are given by
\begin{equation}
    E_{k, n} =\sum_{j \geq 1; j \neq k} \varepsilon_{j, 0} + \varepsilon_{k, n}.
\end{equation}
Similarly, we can define the renormalized case \( E_{k, n}^{(0)} = \varepsilon_{k, n}^{(0)} \). Important to note that
\begin{equation}
    E_{k, 0} = \frac{C}{2} \sum_{j \geq 1} k^{2} = \infty .
\end{equation}
This problem clearly does not exist for the renormalized energy \( E_{k, n}^{(0)} \). From now on, we use the renormalized energy, in a similar approach to the electromagnetic field \cite{Schleich2001}. In the electromagnetic field, this zero-point energy can be observed in an experimental situation as the Casimir effect. 

\subsection{Effects of vacuum: The Phonon Casimir Effect}

The vacuum non-normalized energy is given by

\[
E_{k,0}=\sum_{k}\frac{1}{2}\hbar \omega _{k} ,
\]%
Here we assume an analogous approach of electromagnetic \cite{Schleich2001} field to obtain the action of vacuum on beam, this is the phonon Casimir effect \cite{PhysRevB.103.195434,Motazedifard:17,PhononCasimir1,PhononCasimir2}.
\[
\omega _{k}=\sqrt{\frac{\rho A}{EI}}\left( \frac{k\pi }{L}\right) ^{2}=\tilde{\omega}k^2.
\]%
We can also write it as a function of sound speed, in a solid, it is 
\[
v=\sqrt{\frac{E}{\rho }}
\]%
then 
\begin{eqnarray}
\omega _{k} &=&\frac{1}{v}\sqrt{\frac{A}{I}}\left( \frac{\pi }{L}\right)
^{2}k^{2}.
\end{eqnarray}%
Finally, we can write the zero-point phonon energy as
\[
E_{k,0}=\frac{1}{v}\hbar \sqrt{\frac{A}{I}}\left( \frac{\pi }{L}\right)
^{2}\sum_{k=1}^{\infty }k^{2}.
\]%
If we consider a free system, without $k$ restriction, then 
\begin{equation}
    E_{f}=\frac{1}{v}\hbar \sqrt{\frac{A}{I}}\left( \frac{\pi }{L}\right)
^{2}\int_{0}^{\infty }k^{2}dk.
\end{equation}
The difference $\Delta E=E_{k,0}-E_{f}$ is
\begin{equation}
\Delta E=E_{k,0}-E_{f}=\frac{1}{v}\hbar \sqrt{\frac{A}{I}}\left( \frac{\pi }{L}%
\right) ^{2}\left[ \sum_{k=1}^{\infty }k^{2}-\int_{0}^{\infty }k^{2}dk\right].
\end{equation}

Using a normalizing factor $e^{-\varepsilon k^{2}},$ we have

\begin{eqnarray}
\Delta E &=&E_{0}-E_{f}=\frac{1}{v}\sqrt{\frac{A}{I}}\left( \frac{\pi }{L}\right) ^{2}\hbar
\lim_{M\rightarrow \infty, \varepsilon \rightarrow 0}\left[
\sum_{k=1}^{M}k^{2}e^{-\varepsilon k^{2}}-\int_{0}^{M}k^{2}e^{-\varepsilon
k^{2}}dk\right] \nonumber \\
&=&\frac{1}{v}\sqrt{\frac{A}{I}}\left( \frac{\pi }{L}\right) ^{2}\hbar
\lim_{M\rightarrow \infty, \varepsilon \rightarrow 0}\left\{ - \frac{d}{%
d\varepsilon }\left[ \sum_{k=1}^{M}e^{-\varepsilon
k^{2}}-\int_{0}^{M}e^{-\varepsilon k^{2}}dk\right] \right\} 
\label{eq:DeltaE}
\end{eqnarray}

Observing that

\[
\int_{1}^{M}e^{-\varepsilon k^{2}}dk<\sum_{k=1}^{M}e^{-\varepsilon
k^{2}}<\int_{0}^{M}e^{-\varepsilon k^{2}}dk,
\]
then, for a sufficiently small \( \varepsilon \) we have
\[
\sum_{k=1}^{M}e^{-\varepsilon k^{2}}\approx \frac{1}{2} \left( \int_{1}^{M}e^{-%
\varepsilon k^{2}}dk+\int_{0}^{M}e^{-\varepsilon k^{2}}dk \right)
\]%
or%
\begin{eqnarray}
\sum_{k=1}^{M}e^{-\varepsilon k^{2}}-\int_{0}^{M}e^{-\varepsilon k^{2}}dk
&\approx &-\frac{1}{2} \int_{0}^{1}e^{-\varepsilon k^{2}}dk
\label{eq:sum}
\end{eqnarray}%

Now, inserting \eqref{eq:sum} into \eqref{eq:DeltaE} we have

\begin{eqnarray}
\Delta E &\approx &\frac{1}{v}\sqrt{\frac{A}{I}}\left( \frac{\pi }{L}\right)
^{2}\hbar \lim_{M\rightarrow \infty }\lim_{\varepsilon \rightarrow 0}\left\{ - 
\frac{d}{d\varepsilon }\left[ -\frac{1}{2} \int_{0}^{1}e^{-\varepsilon k^{2}}dk%
\right] \right\}   \\
&=&- \frac{1}{2}\frac{1}{v}\sqrt{\frac{A}{I}}\left( \frac{\pi }{L}\right)
^{2}\hbar \left\{ \int_{0}^{1}k^{2}\,dk\right\}  \\
&=&- \frac{1}{6}\frac{1}{v}\sqrt{\frac{A}{I}}\left( \frac{\pi }{L}\right)
^{2}\hbar.
\end{eqnarray}%
Finally we obtain  
\begin{equation}
\Delta E\approx - \frac{1}{6}\frac{1}{v}\hbar \sqrt{\frac{A}{I}}\left( \frac{\pi }{L}%
\right) ^{2}= - \frac{1}{6}\hbar \sqrt{\frac{\rho A}{EI}}\left( \frac{\pi }{L}%
\right) ^{2}.    
\end{equation}
Now the energy per area ($\mathcal{V}$) acting in the beam cross section is

\begin{equation}
\mathcal{V}\approx - \frac{1}{6}\frac{1}{v}\hbar \sqrt{\frac{1}{IA}}\left( \frac{\pi }{L}%
\right) ^{2}= - \frac{1}{6}\hbar \sqrt{\frac{\rho }{EIA}}\left( \frac{\pi }{L}%
\right) ^{2},    
\end{equation}
and thus the force per unit area $F=-\frac{\partial \mathcal{V}}{\partial L},$ is 
\begin{equation}
    F\approx - \frac{1}{3}\hbar \sqrt{\frac{\rho }{EIA}}\frac{\pi ^{2}}{L^{3}}= - \frac{1}{3}\frac{1}{AL}\hbar\tilde{\omega}.
    \label{eq:Casimir_Force}
\end{equation}
Note from \eqref{eq:Casimir_Force} that $F\propto L^{-3}$, and it is proportional to the energy fundamental state of the first mode, $k=1$, divided by the volume of the beam. A discussion of a similar phonon Casimir effect can be found in references \cite{PhysRevB.103.195434,Motazedifard:17}. Consequently, it is very small and attractive. In the next calculations, we will neglect this force. On the other hand, if the term $AL$ is small and $F$ isn't negligible, the force $F$ should be considered from the beginning in  the Euler-Bernoulli modeling \cite{Cleiser}.

\subsection{Degeneracy and effects of boundary conditions} \label{boundary}

In this subsection, we investigate if this model presents energy degeneracy, i.e. for which values for the pairs of indices \( (k,n) \neq (k',n') \) we could have \( E_{k,n} = E_{k', n'} \). First, we observe that 
\begin{align*}
    E_{k,n} - E_{k', n'} &= \left( \sum_{j \neq k} \varepsilon_{j,0} + \varepsilon_{k, n} \right) - \left( \sum_{j \neq k'} \varepsilon_{j,0} + \varepsilon_{k', n'} \right) = \\
    &= (\varepsilon_{k, n} + \varepsilon_{k', 0}) - (\varepsilon_{k', n'} + \varepsilon_{k, 0}) = \\
    &= (\varepsilon_{k, n} - \varepsilon_{k, 0}) - (\varepsilon_{k', n'} - \varepsilon_{k', 0}) = \\
    &= \varepsilon_{k, n}^{(0)} - \varepsilon_{k', n'}^{(0)} = E_{k,n}^{(0)} - E_{k', n'}^{(0)}.
\end{align*}

Therefore we have  \( E_{k,n} = E_{k', n'} \) if, and  only if, \( \varepsilon_{k, n}^{(0)} = \varepsilon_{k', n'}^{(0)} \), and consequently 
\begin{equation}
    C k^{2} n = C (k')^{2} n' \Rightarrow n = \left( \frac{k'}{k} \right)^{2} n'.
\end{equation}
For instance, if  \( k' = 2k \), then  \( n = 4n' \), a particular case is  \( E_{1,4} = E_{2,1} \).

Experimentally, the boundary conditions have an impact on the quality factor of the resonator $Q^{-1}$. Takamura and collaborators \cite{mi7090158} have observed different $Q^{-1}$ for free-free edges and  doubly clamped graphene resonators. A dependence of $Q^{-1}$ on the boundary conditions has also been found in classical molecular dynamics simulations \cite{C2NR30493G}. In reference \cite{10.1063/1.3611396}, they found a fitting for the decay rate as $e^{-\eta\omega t}$, where $\eta$ is a constant and $\omega$ is the natural beam frequency.

In our model, the  boundary determines $\lambda_k$ and consequently $\omega_0$; the value of $\lambda_k$ given by \ref{eq:classicbeam_normalmodes} is found for the hinged-hinged boundary condition; for other boundary conditions, it does not have such a simple analytical formula. The degeneracy, found in the previous section, only exists because the ratio between two consecutive values of $\lambda_k$ is a rational number. Let us now analyze other possibilities that can be found in real situations, namely, clamped-clamped, clamped-hinged, clamped-free, and free-free. The solution \eqref{eq:classicbeam_normalmodes} is obtained by the roots of the characteristic equation $F_0(\lambda_k L)=\sin (\lambda_k L)$; for other boundary conditions, we have different characteristic equations given in table \ref{TAB1}.

\begin{figure}
    \centering
    \includegraphics[width=0.8\linewidth]{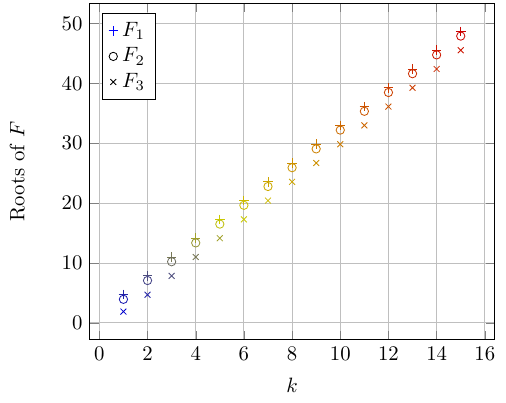}
    \caption{Roots of $F_{j}$ as function of $k$ for $L=1$.}
    \label{fig:roots-of-F}
\end{figure}

\begin{figure}
    \centering
    \includegraphics[width=0.8\linewidth]{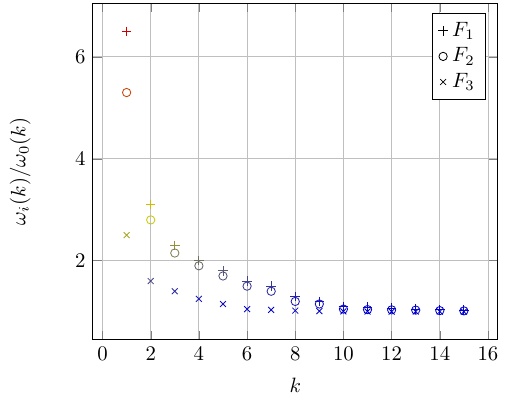}
    \caption{Ratio of $\frac{\omega_i}{\omega_0}$ as function of $k$, and $\omega_0(k)=\frac{ \pi^{2}}{L^{2}} \sqrt{\frac{\rho A}{EI}} \, k^{2}$.}
    \label{fig:omega_i}
\end{figure}

\begin{table}[ht]
    \centering
    \begin{tabular}{|c|c|}
    \hline
       clamped-clamped  &  $F_1(x)=\cos{x}\cosh{x}-1$\\
       \hline
        clamped-hinged & $F_2(x)=\cos{x}\sinh{x}-\sin{x}\cosh{x}$ \\
        \hline
        clamped-free &  $F_3(x)=\cos{x}\cosh{x}+1$\\
        \hline
        free-free &  $F_4(x)=\cos{x}\cosh{x}-1=F_1(x)$\\
        \hline
    \end{tabular}
    \caption{Characteristic Equation for determining $\lambda_k$ for different boundary conditions, and $x=\lambda_k L$. }
    \label{TAB1}
\end{table}
In figure \ref{fig:roots-of-F} we see the first roots of $F_1$, $F_2$ and $F_3$. They differ little from the roots of $F_0$, the hinged-hinged case, which is evident in figure \ref{fig:omega_i} where $\omega_0$ refers to the hinged-hinged case.
Considering the functions $F_j$, for $j\neq 0$, we have for small values of $k$ the difference $\lambda_{k+1}-\lambda_k$ has no simple relation to $\pi$, this for all functions $F_m(x)$, and $m=1,...,4$. For $k>4$ we have that  $\frac{\lambda_{k+1}-\lambda_k}{\pi L} \approx 1$, and $\lambda_k L\approx q\pi$ where $q\in \mathbb{Q}$. For $k>10$ we have that $\frac{\lambda_{k+1}-\lambda_k}{\pi L} = 1+\epsilon$ and $|\epsilon| \leq 10^{-13}$.

Thus, for larger values of $k$, there will always be a quasi-degeneracy, which in an experiment may be difficult to differentiate from a real degeneracy.

\section{Decoherence ``free'' subspace} \label{sec:Dechoerence_free}

\begin{figure}
    \centering
    \includegraphics[width=0.5\linewidth]{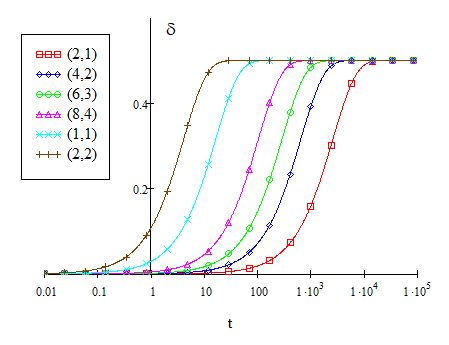}
    \caption{Time evolution of linear entropy of the state \ref{psi_D} for $a=b=\frac{1}{\sqrt{2}}$, $j=1$ and $k=2$. The values of $(m,n)$ are given in the legend. The others constants are chosen as $\frac{\rho A}{EI}=1$, $\hbar=1$, and  $\Lambda=\frac{1}{(10\pi)^2}$. The beam contour condition is clamped-hinged.}
    \label{fig:linear entropy}
\end{figure}

Now we consider that the beam is not isolated, but it is connected by its surroundings interacting by inelastic collisions, or, in other words, there is no energy exchange. This is the phase-dumping reservoir \cite{Phasereservoir}. We consider a reservoir model with a large number of harmonic oscillators. On the other hand, the effectiveness of the phase dumping reservoir depends solely on the effective Hilbert space \cite{HilbertSpace}. The Hamiltonian for the total system environment is given by
\begin{equation}
    \hat{H}=\hat{H}_S+\hat{H}_I+\hat{H}_E,
\end{equation}
where 
\begin{equation}
    \hat{H}_S=\sum_k \hat{\mathcal{H}}_k,
\end{equation}

\begin{equation}
    \hat{H}_I=\lambda \sum_{k,j} \hat{\mathcal{H}}_k b^{\dagger}_jb_j,
\end{equation}
and
\begin{equation}
    \hat{H}_E= \sum_{j} \omega_j b^{\dagger}_jb_j.
\end{equation}
Here $\lambda$ is a coupling constant, $b_k$ ($b^\dagger$) is the destruction (creation) operator of the $k$-th environment oscillator, and $\omega$ is its frequency. We will consider only two modes, then $\hat{H}_S= \hat{\mathcal{H}}_j+\hat{\mathcal{H}}_k$. The eigenstates of  $\hat{H}_S$ are now of the form
\begin{equation}
    \hat{H}_S\ket{\phi_{m,n}}= \hat{H}_S\ket{m,n}=E_{m,n}\ket{m,n},
\end{equation}
where $E_{m,n}=\hbar\omega_j m+\hbar\omega_k n$.
The time evolution of states in this basis can be obtained by 
\begin{equation}
    \ket{\phi_{m,n}}\bra{\phi_{m',n'}}(t)= e^{-\left[ \Lambda(\overline{n})(\Delta E)^2 \right]t}\ket{\phi_{m,n}}\bra{\phi_{m',n'}},
    \label{rho_nm}
\end{equation}
where $\Delta E=E_{m,n}-E_{m',n'}$ and $\Lambda(\overline{n})$ is a function of the mean photon number of the environment $\overline{n}$, and the coupling constant $\lambda$, see \cite{Phasereservoir,HilbertSpace} for details. The diagonal states are not affected by the phase dumping reservoir, since $\Delta E=0$, but not only the diagonal states. Let's consider the state
\begin{equation}
    \ket{\psi_D}=a\ket{0}_j\otimes \ket{n}_k+b\ket{m}_j\otimes\ket{0}_k,
    \label{psi_D}
\end{equation}
and we assume that the state $\ket{0}_j\otimes \ket{n}_k$ and $\ket{m}_j\otimes \ket{0}_k$ are  degenerate, then $ \ket{\psi_D}$ is preserved. 
If the contour condition is not hinged-hinged, then there is no degeneracy and the state is not preserved, but, as previously discussed, there are some pairs that are nearly degenerated. In this condition, the state is not preserved, but it has a greater lifetime. 
In figure \ref{fig:linear entropy}, we plot the linear entropy $\delta$ that is defined as $\delta=1-\Tr{\rho^2}$ for the state \ref{psi_D} in a clamped-hinged contour condition. As we can see, there is a large variation in the decoherence time. In a phase-damping reservoir, the decoherence time usually depends on the difference in the energy levels. Here, the difference lies in the fact that the energy levels are not equally spaced if we consider two or more modes. The natural choice would be $(j,m)=(1,1)$ and $(k,n)=(2,1)$, but the biggest decoherence time was found for $(j,m)=(1,2)$ and $(k,n)=(2,1)$. For larger values of $k$ we have that $\omega_k\approx\omega_0(1) (k+\frac{1}{4})^2$. The factor $\frac{1}{4}$ in $\omega_k$ drives the ``degeneracy'' to larger values of $m$ and $n$, although the energy difference becomes proportionally smaller; in fact, it is not small, and those states are not less affected by the environment.
\subsubsection{Some considerations about a more realistic thermal bath}
If we consider a zero-temperature thermal bath \cite{BOSCODEMAGALHAES2004}, we have a decay rate that depends on $D(\omega)d\omega$, which represents the number of modes in the frequency interval between $\omega$ and $\omega+d\omega$ of the thermal bath. Since the modes represent vibrational states of the same beam, the density of states of the thermal bath is the same. However, since different vibration modes represent deformations with different wavelengths, it is reasonable to imagine that the coupling between the baths must be different for each vibrational mode, and therefore the rates of loss of coherence ($\kappa_i$) can be very different, where $\kappa_i$ is the decay rate of the vibrational mode $i$. If we assume that the decay rate is $\kappa=\eta\omega$ as shown in ref. \cite{10.1063/1.3611396}, then we can have  $\kappa_A\ll\kappa_B$. If the modes were uncoupled, the mode $B$ would lose its coherence much faster than mode $A$. Now, including a coupling, it makes them decay at a rate $\kappa_{AB}$, and we have $\kappa_{AB}\approx\kappa_A$, which is a necessary condition for the existence of a decoherence-free subspace \cite{BOSCODEMAGALHAES2004}. In this case, it is not a subspace without decoherence but one with a significantly longer half-life.

\section{Conclusions} \label{sec:Conclusions}
As we can see, the semiclassical treatment of the Euler-Bernoulli nanobeam has many distinct facets. As we consider the existence of infinite vibrational modes, we are forced to perform a renormalization. Treating this renormalization in a way analogous to the quantization of the electromagnetic field, we have a thermal, or phonon Casimir effect. Also, as a consequence of the infinite vibrational modes, we observe that, depending on the boundary condition, there may be a degeneracy in the Hamiltonian spectrum. The degeneracy ends up having other consequences. For dispersive baths, we observe that there are decoherence-free subspaces; these spaces still exist approximately even when we change the boundary conditions; these are the quasi-degenerate states. For these cases, the decoherence time may be orders of magnitude shorter, so they are possible candidates for use in quantum computing applications.
\begin{acknowledgments}
This work was supported by the Fundação de Amparo à Pesquisa do
Estado de Minas Gerais (FAPEMIG) ; the
Coordenação de Aperfeiçoamento de Pessoal de Nível Superior (CAPES); and the National Council for Scientific and Technological
Development (CNPq).
\end{acknowledgments}

\bibliography{apssamp}

\end{document}